# Topological Splitting of Sondheimer Oscillations in Thin-Film Weyl Semimetals

Toshihito Osada*

*Department of Applied Physics, The University of Tokyo,*

*7-3-1 Hongo, Bunkyo-ku, Tokyo 113-8656, Japan.*

We discuss topological effects on Sondheimer oscillations in the magnetoresistance of thin-film Weyl semimetals. The Sondheimer oscillations are periodic in magnetic field and originate from the commensurability between the film thickness and the cyclotron-orbit pitch. Using Boltzmann transport theory, we investigate the effects of Berry curvature and orbital magnetic moment on the oscillations. We find that the Sondheimer oscillations split into two branches with different frequencies and that each branch becomes chirped because its frequency depends on magnetic field. These features originate from the orbital-Zeeman-induced deformation of the Fermi surface. The same mechanism also applies to Dirac semimetals.

Sondheimer oscillations (SOs) are magnetic-field-periodic oscillations in the transverse magnetoresistance of thin-film or plate-like crystals [1,2]. They arise from the commensurability between the sample thickness and the pitch of spiral electron motion in a magnetic field [Fig. 1(a)] and provide information on the Fermi surface near its limiting points. Following Sondheimer's semiclassical Boltzmann description of the thin-film size effect [1], SOs were observed in several metals and used to probe their Fermi surfaces [3,4]. More recently, related phenomena have attracted renewed interest in topological materials and thin-film devices [5–7]. SO-like oscillations have also been observed in the quantum-limit regime of graphite [8,9], motivating quantum-transport theories that connect conventional and quantum SOs [10,11]; the latter can also be sensitive to the Berry phase in two-dimensional zero-gap systems [11].

In this Letter, we investigate conventional SOs in thin-film Weyl semimetals. Weyl nodes act as sources or sinks of Berry curvature and are accompanied by a finite orbital magnetic moment, which is directly related to the Berry curvature in two-band systems [12]. In a magnetic field, the orbital magnetic moment modifies the band energy through the orbital Zeeman coupling. We examine how these topological corrections modify the SOs.

We first consider the semiclassical orbital motion of an electron in a single isotropic Weyl cone described by $\hat{H} = \chi\hbar v_{\mathrm{F}}\mathbf{k}\cdot\boldsymbol{\sigma}$, where $\chi = \pm 1$ denotes the chirality, $v_{\mathrm{F}}$ is the constant velocity, and $\boldsymbol{\sigma}$ is the vector of Pauli matrices. For the conduction band, the energy, Berry curvature, and orbital magnetic moment are given by $E_{\mathbf{k}} = \hbar v_{\mathrm{F}} k$, $\boldsymbol{\Omega}_{\mathbf{k}} = -\chi\mathbf{n}_{\mathbf{k}}/(2k^2)$, and $\mathbf{m}_{\mathbf{k}} = (e/\hbar)E_{\mathbf{k}}\boldsymbol{\Omega}_{\mathbf{k}} = -\chi e v_{\mathrm{F}}\mathbf{n}_{\mathbf{k}}/(2k)$, where $k = |\mathbf{k}|$ and $\mathbf{n}_{\mathbf{k}} = \mathbf{k}/k$. We consider a film of thickness $L_z$ along the $z$-direction and apply a magnetic field $\mathbf{B} = (0,\ 0,\ B)$. The semiclassical equations of motion, $\dot{\mathbf{r}} = \tilde{\mathbf{v}}_{\mathbf{k}} - \dot{\mathbf{k}} \times \mathbf{\Omega}_{\mathbf{k}}$, $\hbar\dot{\mathbf{k}} = -e\dot{\mathbf{r}} \times \mathbf{B}$ can be rewritten as Eqs. (1a) and (1b).

$$\dot{\mathbf{r}} = \left(1 + \frac{e}{\hbar}\mathbf{B} \cdot \mathbf{\Omega}_{\mathbf{k}}\right)^{-1} \left\{\tilde{\mathbf{v}}_{\mathbf{k}} + \frac{e}{\hbar}(\mathbf{\Omega}_{\mathbf{k}} \cdot \tilde{\mathbf{v}}_{\mathbf{k}})\mathbf{B}\right\}, \tag{1a}$$

$$\dot{\mathbf{k}} = \left(1 + \frac{e}{\hbar}\mathbf{B} \cdot \mathbf{\Omega}_{\mathbf{k}}\right)^{-1} \left(-\frac{e}{\hbar}\tilde{\mathbf{v}}_{\mathbf{k}} \times \mathbf{B}\right). \tag{1b}$$

Here, $\tilde{\mathbf{v}}_{\mathbf{k}} = (1/\hbar)\partial\tilde{E}_{\mathbf{k}} / \partial\mathbf{k}$ is the group velocity of the field-modified band, $\tilde{E}_{\mathbf{k}} = E_{\mathbf{k}} - \mathbf{m}_{\mathbf{k}} \cdot \mathbf{B}$, where the second term represents the orbital Zeeman energy. Equation (1b) shows that the **k**-space orbit follows the intersection of a constant-$\tilde{E}_{\mathbf{k}}$ surface with a plane perpendicular to the magnetic field.

We represent the wave vector as $\mathbf{k} = k(\sqrt{1-u^2}\cos\varphi,\ \sqrt{1-u^2}\sin\varphi,\ u)$, $u = \cos\theta$, and introduce the dimensionless parameter $b_k \equiv eB/(2\hbar k^2) = (e/\hbar)B\left|\mathbf{\Omega}_{\mathbf{k}}\right|$, which characterizes the strength of the topological correction. In the following, we retain terms up to first order in $b_k$. The invariant measure becomes $1 + (e/\hbar)\mathbf{B} \cdot \mathbf{\Omega}_{\mathbf{k}} = 1 - \chi b_k u$. The field-modified group velocity is $\tilde{\mathbf{v}}_{\mathbf{k}} = v_{\mathrm{F}}\left(1 - 2\chi b_k u\right)\mathbf{n}_{\mathbf{k}} + \chi v_{\mathrm{F}} b_k \mathbf{n}_z$, and Eq. (1a) then gives, to first order in $b_k$, $\dot{\mathbf{r}} \simeq v_{\mathrm{F}}(1 - \chi b_k u)\mathbf{n}_{\mathbf{k}}$. Thus, the total velocity $\dot{\mathbf{r}}$ remains parallel to **k**. Its $z$-component is $\dot{z} = \left|\dot{\mathbf{r}}\right|\cos\theta = v_{\mathrm{F}} u - \chi v_{\mathrm{F}} b_k u^2$, so that the topological correction has the same sign for $u > 0$ and $u < 0$. This velocity asymmetry is related to the semiclassical motion underlying the chiral magnetic effect. The velocity field for $\chi = +1$ is illustrated in Fig. 1(b).

Equation (1b) then yields a circular orbit at fixed $k = k_0$ and $u = u_0$, $k_x = k_0\sqrt{1-u_0^{\,2}}\cos(\tilde{\omega}_c t + \varphi_0)$, $k_y = k_0\sqrt{1-u_0^{\,2}}\sin(\tilde{\omega}_c t + \varphi_0)$, $k_z = k_0 u_0$, with $\tilde{\omega}_c = (eBv_F / \hbar k_0)(1 - \chi b_{k_0} u_0)$. Here, $k_0$, $\varphi_0$, and $u_0$ specify the orbit at $t$=0.

We assume that electrons are doped into the conduction band and define the zero-field Fermi energy as $E_F = \hbar v_F k_F > 0$, where $k_F$ is the Fermi wave number at $B$ = 0. In a magnetic field, the Fermi surface (FS), defined by $\tilde{E}_{\mathbf{k}} = E_F$, is deformed by the orbital Zeeman effect. Since $\tilde{E}_{\mathbf{k}} = \hbar v_F k(1 + \chi b_k u)$, the FS is given, to first order in $b_k$, by $k = k_F (1 + \chi b_k u)^{-1} \simeq k_F (1 - \chi b_k u)$. The deformed FS is illustrated in Fig. 1(b). For an orbit on this modified FS, $k_0 \simeq k_F (1 - \chi b_{k_F} u_0)$. Substituting this into the cyclotron frequency gives $\tilde{\omega}_c = (eBv_F / \hbar k_0)(1 - \chi b_{k_0} u_0) \simeq eBv_F / \hbar k_F \equiv \omega_{cF}$. Thus, the first-order topological correction to the cyclotron frequency cancels on the modified FS. Accordingly, $b_k$ can be replaced by $b_{k_F}$ to first order in $b_k$.

We next consider the real-space motion in a film of finite thickness $0 \leqq z \leqq L_z$. Using $\dot{\mathbf{r}} \simeq v_F (1 - \chi b_{k_F} u_0) \mathbf{k} / k_0$, the velocity components are

$$\dot{x} = v_F (1 - \chi b_{k_F} u_0) \sqrt{1 - u_0{}^2} \cos(\omega_{cF} t + \varphi_0), \tag{2a}$$

$$\dot{y} = v_F (1 - \chi b_{k_F} u_0) \sqrt{1 - u_0{}^2} \sin(\omega_{cF} t + \varphi_0), \tag{2b}$$

$$\dot{z} = v_F (1 - \chi b_{k_F} u_0) u_0 . \tag{2c}$$

The $z$-coordinate is therefore $z(t) = z_0 + v_F (1 - \chi b_{k_F} u_0) u_0 t$. We define the survival probability $P(t; \mathbf{k}_0)$ as the fraction of initial positions $z_0$ for which the trajectory, traced backward to past time $t < 0$, does not reach either surface at $z$ = 0 or $L_z$. It is the proportion of $z_0$ for which the displacement $|z - z_0|$ over the time interval $|t|$ falls within the thickness $L_z$. Since $|z - z_0| = v_F (1 - \chi b_{k_F} u_0) |u_0| |t|$, the allowed range of $z_0$ has width $L_z - |z - z_0|$,

giving

$$P(t;\mathbf{k}_0)=\max\left[0,\ 1+\lambda\left(1-\chi b_{k_{\rm F}}u_0\right)\left|u_0\right|\frac{t}{\tau}\right], \tag{3}$$

where $t<0$ and $\lambda=v_{\rm F}\tau/L_z$ characterizes the strength of the finite-size effect.

We evaluate the in-plane conductivities $\sigma_{xx}$ and $\sigma_{xy}$ using the semiclassical Boltzmann transport theory [13]. We assume a constant bulk relaxation time $\tau$ and completely diffusive surface scattering, which erases the momentum memory at the boundaries. Under these assumptions, the conductivity is given by the following Chambers formula.

$$\sigma_{\mu\nu}=\frac{e^2}{(2\pi)^3}\iiint \dot{r}_\mu(\mathbf{k}_0)\langle\dot{r}_\nu\rangle\left(-\frac{\partial f^0}{\partial\tilde{E}_{\mathbf{k}_0}}\right)\left(1+\frac{e}{\hbar}\mathbf{B}\cdot\mathbf{\Omega}_{\mathbf{k}}\right)d\mathbf{k}_0{}^3, \tag{4a}$$

$$\langle\dot{r}_\nu\rangle=\int_{-\infty}^{0}\dot{r}_\nu\left(\mathbf{k}(t;\mathbf{k}_0)\right)e^{\frac{t}{\tau}}P(t;\mathbf{k}_0)dt. \tag{4b}$$

The factor $1+(e/\hbar)\mathbf{B}\cdot\mathbf{\Omega}_{\mathbf{k}}$ is the Berry-curvature-modified **k**-space measure. Using Eqs. (2) and (3), we write $\sigma_{xx}=\bar{\sigma}_{xx}+\delta\sigma_{xx}$, $\sigma_{xy}=\bar{\sigma}_{xy}+\delta\sigma_{xy}$, where barred and $\delta$ terms denote the background and oscillatory contributions, respectively.

$$\bar{\sigma}_{xx}=\frac{3}{4}\sigma_0\int_{-1}^{1}(1-b_{k_{\rm F}}u)^4(1-u^2)\left[\frac{1}{1+(\omega_{c\rm F}\tau)^2}+a(u)\frac{(\omega_{c\rm F}\tau)^2-1}{\{1+(\omega_{c\rm F}\tau)^2\}^2}\right]du, \tag{5a}$$

$$\delta\sigma_{xx}=\frac{3}{4}\sigma_0\int_{-1}^{1}(1-b_{k_{\rm F}}u)^4(1-u^2)\frac{a(u)e^{-1/a(u)}}{1+(\omega_{c\rm F}\tau)^2}\cos\left(\frac{\omega_{c\rm F}\tau}{a(u)}+2\tan^{-1}\omega_{c\rm F}\tau\right)du, \tag{5b}$$

$$\bar{\sigma}_{xy}=\frac{3}{4}\sigma_0\int_{-1}^{1}(1-b_{k_{\rm F}}u)^4(1-u^2)\left[\frac{-\omega_{c\rm F}\tau}{1+(\omega_{c\rm F}\tau)^2}+2a(u)\frac{\omega_{c\rm F}\tau}{\{1+(\omega_{c\rm F}\tau)^2\}^2}\right]du, \tag{6a}$$

$$\delta\sigma_{xy}=-\frac{3}{4}\sigma_0\int_{-1}^{1}(1-b_{k_{\rm F}}u)^4(1-u^2)\frac{a(u)e^{-1/a(u)}}{1+(\omega_{c\rm F}\tau)^2}\sin\left(\frac{\omega_{c\rm F}\tau}{a(u)}+2\tan^{-1}\omega_{c\rm F}\tau\right)du. \tag{6b}$$

Here, $\sigma_0 = e^2 k_F{}^2 v_F \tau / (6\pi^2\hbar)$ is the Drude conductivity of a Weyl cone, and $a(u) = \lambda |u| (1 - b_{k_F} u)$. We have absorbed the chirality $\chi$ into the integration variable, $u = \chi u_0$, so that the conductivity does not depend on the chirality of the Weyl cone. In the following, all amplitudes and background terms are retained up to first order in $b_{k_F}$, while the resulting first-order phase correction is kept unexpanded in the oscillatory factors.

The longitudinal resistivity is obtained from $\rho_{xx} = \sigma_{xx} / (\sigma_{xx}{}^2 + \sigma_{xy}{}^2)$. We define the background resistivity by $\bar{\rho}_{xx} = \bar{\sigma}_{xx} / (\bar{\sigma}_{xx}{}^2 + \bar{\sigma}_{xy}{}^2)$, and the oscillatory part by $\delta\rho_{xx} = \rho_{xx} - \bar{\rho}_{xx}$. For $N$ Weyl cones with the same $k_F$, $v_F$, and $\tau$, the total conductivity is obtained by replacing $\sigma_0$ with $N\sigma_0$.

Expanding the background term (5a) to first order in $b_{k_F}$, the odd-in-$u$ correction vanishes after integration over $-1 \leqq u \leqq 1$, yielding

$$\bar{\sigma}_{xx} \simeq \frac{\sigma_0}{1 + (\omega_{cF}\tau)^2} \left\{ 1 + \frac{3}{8} \lambda \frac{(\omega_{cF}\tau)^2 - 1}{(\omega_{cF}\tau)^2 + 1} \right\}. \tag{7a}$$

To extract the leading oscillatory contribution in a high-field expansion in $\lambda / (\omega_{cF}\tau) << 1$, we integrate Eq. (5b) by parts twice and retain the boundary terms at $u = \pm 1$, which correspond to the limiting points.

$$\delta\sigma_{xx} \sim \frac{3}{2} \lambda \frac{\sigma_0}{1 + (\omega_{cF}\tau)^2} e^{-\frac{1}{\lambda}} \left( \frac{\lambda}{\omega_{cF}\tau} \right)^2 \sum_{\pm} \left\{ 1 \mp \left( 5 + \frac{1}{\lambda} \right) b_{k_F} \right\} \cos\left( (1 \pm b_{k_F}) \frac{\omega_{cF}\tau}{\lambda} - \frac{2}{\omega_{cF}\tau} \right), \tag{7b}$$

At sufficiently high fields, $\omega_{cF}\tau >> 1$, $|\bar{\sigma}_{xy}| >> \bar{\sigma}_{xx}$ and $|\delta\bar{\sigma}_{xx}| \sim |\delta\bar{\sigma}_{xy}|$, so that $\delta\rho_{xx} / \bar{\rho}_{xx} \simeq \delta\sigma_{xx} / \bar{\sigma}_{xx}$.

$$\frac{\delta\rho_{xx}}{\bar{\rho}_{xx}} \sim \frac{12\lambda}{8 + 3\lambda} \frac{e^{-1/\lambda}}{\beta^2} \sum_{\pm} \left\{ 1 \mp \left( 5 + \frac{1}{\lambda} \right) \frac{\beta}{2k_F L_z} \right\} \cos\left( \beta \pm \frac{\beta^2}{2k_F L_z} - \frac{2}{\lambda\beta} \right). \tag{8}$$

Here, $\beta = eBL_z / (\hbar k_F) = \omega_{cF}\tau / \lambda = 2(k_F L_z) b_{k_F}$ is a dimensionless magnetic field. Since $b_{k_F} = \beta / (2k_F L_z)$, the oscillation phase of two branches is $\Phi_\pm(\beta) = \beta \pm \beta^2 / (2k_F L_z) - 2 / (\lambda\beta)$. Therefore, the corresponding local frequencies are

$$\omega_\beta^\pm = \frac{d\Phi_\pm}{d\beta} \simeq 1 \pm \frac{\beta}{k_F L_z}, \tag{9}$$

where the last term $2 / (\lambda\beta^2)$ has been neglected at high fields. The local oscillation period $\Delta\beta_\pm$ is obtained as $\Delta\beta_\pm = 2\pi / \omega_\beta^\pm \simeq 2\pi / \{1 \pm \beta / (k_F L_z)\}$.

For $b_{k_F} = 0$, the oscillation is strictly periodic in $\beta$ with period $\Delta\beta = 2\pi$, corresponding to $\Delta B = 2\pi\hbar k_F / (eL_z)$. This is the conventional SO period determined by the commensurability between $L_z$ and the cyclotron-orbit pitch near the limiting points $u = \pm 1$ on the spherical FS. In contrast, finite $b_{k_F}$ in Weyl semimetals splits the SO into two branches whose frequency separation increases linearly with magnetic field. Because the local frequencies themselves vary with $\beta$, each branch is chirped. The two branches also acquire different amplitudes, with the lower-frequency branch becoming stronger.

The splitting originates from the orbital-Zeeman-induced FS deformation. On the modified FS, the velocity along $z$ is $\dot{z} \simeq v_F(1 - \chi b_{k_F} u_0) u_0$, whereas the cyclotron frequency remains $\omega_{cF}$ to first order. The pitch of the modified orbit $|\dot{z}| \cdot (2\pi / \omega_{cF}) = \{2\pi\hbar k_F / (eB)\} |u_0| (1 - \chi b_{k_F} u_0)$, which depends on $u_0$. Its commensurability with the film thickness $L_z$ gives the local period $\Delta\beta = 2\pi |u_0| / \{1 + \chi u_0 \beta / (k_F L_z)\}$. At limiting points $u_0 = \pm 1$, the SO periods become $\Delta\beta = 2\pi / \{1 \pm \chi\beta / (k_F L_z)\}$, in agreement with Eq. (9), since the two limiting points correspond to $\chi u_0 = \pm 1$. Therefore, the two split SO branches originate from the contribution from each of the limiting points of the FS deformed by the orbital Zeeman effect.

We next examine these features numerically. Figure 2(a) shows the magnetic-field dependence of the resistivity $\rho_{xx}$ for a topologically trivial metal ($b_{k_{\rm F}} = 0$) for several values of the size parameter $\lambda$. The resistivity is calculated from Eqs. (5) and (6) and plotted as a function of the dimensionless field $\beta = eBL_z / (\hbar k_{\rm F})$. At high fields, the background resistivity approaches $\bar{\rho}_{xx} / \sigma_0^{-1} = 1 + 3\lambda / 8$, showing that the resistance increases with decreasing film thickness. This is caused by the increase of the surface scattering contribution. Superimposed on this background are the conventional SOs, which are periodic in magnetic field and whose amplitude decreases approximately as $B^{-2}$.

Figure 2(b) shows the oscillatory component $\delta\rho_{xx}$ for Weyl semimetals with fixed $\lambda = 1$ and several values of $k_{\rm F}L_z$. The topological correction, $b_{k_{\rm F}} = \beta / (2k_{\rm F}L_z)$, becomes stronger as $k_{\rm F}$ decreases towards the Weyl point. In the numerical calculation, the background and amplitude factors are retained to first order in $b_{k_{\rm F}}$, while the resulting first-order phase correction is kept unexpanded. The curves are vertically offset for clarity. For $k_{\rm F}L_z = 10000$, the topological correction is negligible and the conventional SO is recovered. With decreasing $k_{\rm F}L_z$, the oscillation becomes progressively modulated, exhibiting a characteristic beating structure.

To characterize the modification of the SO more quantitatively, we calculate the Fourier power spectra of the data in Fig. 2(b) over the range $10 \leqq \beta \leqq 100$, as shown in Fig. 3(a). For $k_{\rm F}L_z = 10000$, the spectrum has a single peak at $\omega_\beta = 2\pi / \Delta\beta = 1$, corresponding to the conventional SO with $\Delta\beta = 2\pi$. As $k_{\rm F}L_z$ decreases, the peak splits into two branches and broadens. The increasing separation of the two peaks reflects the topological splitting, while the broadening indicates the deviation of each branch from perfect periodicity. The low-frequency peak also becomes more pronounced, consistent with the analytical prediction for the branch amplitudes.

To resolve the field dependence of the split frequencies, we further calculate moving-window Fourier power spectra for $k_F L_z = 200$, as shown in Fig. 3(b). A Hann window of width $\Delta\beta = 30$ is used, and the window center is shifted along the $\beta$ axis. As the window moves to higher $\beta$, the separation between the two peaks increases. This directly demonstrates that the local frequencies of the two SO branches vary with magnetic field, i.e., that both branches are chirped, in agreement with Eq. (9).

Finally, we estimate the expected SO modification for realistic Weyl semimetals. In typical materials such as TaAs or NbP, the carrier density, Fermi velocity, and scattering time are of order $n = 10^{18}-10^{20}$ cm$^{-3}$, $v_F = (1-8) \times 10^5$ m/s, and $\tau = 1-100$ ps [12]. Taking $k_F = 0.4$ nm$^{-1}$, $v_F = 2 \times 10^5$ m/s, $\tau = 2.5$ ps, and $L_z = 500$ nm, we obtain $n = 1.08 \times 10^{18}$ cm$^{-3}$ (per cone), $k_F L_z = 200$, $b_{kF} = 0.047$, $\lambda = 1.0$, and $\omega_{cF}\tau = 19.0$ at $B = 10$ T. The corresponding local SO periods are $\Delta B_{\pm} = 3.3/(1 \pm 0.094)$ T, indicating that the splitting should be experimentally observable. Observation of the split SO would provide a direct probe of the orbital-Zeeman-induced Fermi-surface deformation.

The present results are also applicable to Dirac semimetals, where Weyl cones of opposite chirality are degenerate. Since the conductivity obtained here is independent of chirality, the splitting and chirping of the Sondheimer oscillations are expected to survive in Dirac semimetals as well.

In conclusion, we have investigated SOs in thin-film Weyl semimetals within semiclassical Boltzmann transport theory including Berry-curvature and orbital-magnetic-moment effects. We find that the oscillations split into two branches and that each branch is chirped. These features originate from the inequivalence of the two limiting points of the Fermi surface deformed by the orbital Zeeman effect. Observation of the split and chirped SOs would therefore provide a probe of this field-induced Fermi-surface deformation. Since the conductivity is independent of chirality in the present model, the same mechanism also applies to Dirac semimetals with degenerate cones of opposite chirality.

**Acknowledgements**

This work was partially supported by JSPS KAKENHI Grant Numbers JP26K06975 and JP26H00589. The author thanks Prof. Tatsuo Hasegawa and Prof. Yasuhiro H. Matsuda for providing an excellent research environment.

## References

*osada@issp.u-tokyo.ac.jp

**Figure 1** (Osada)

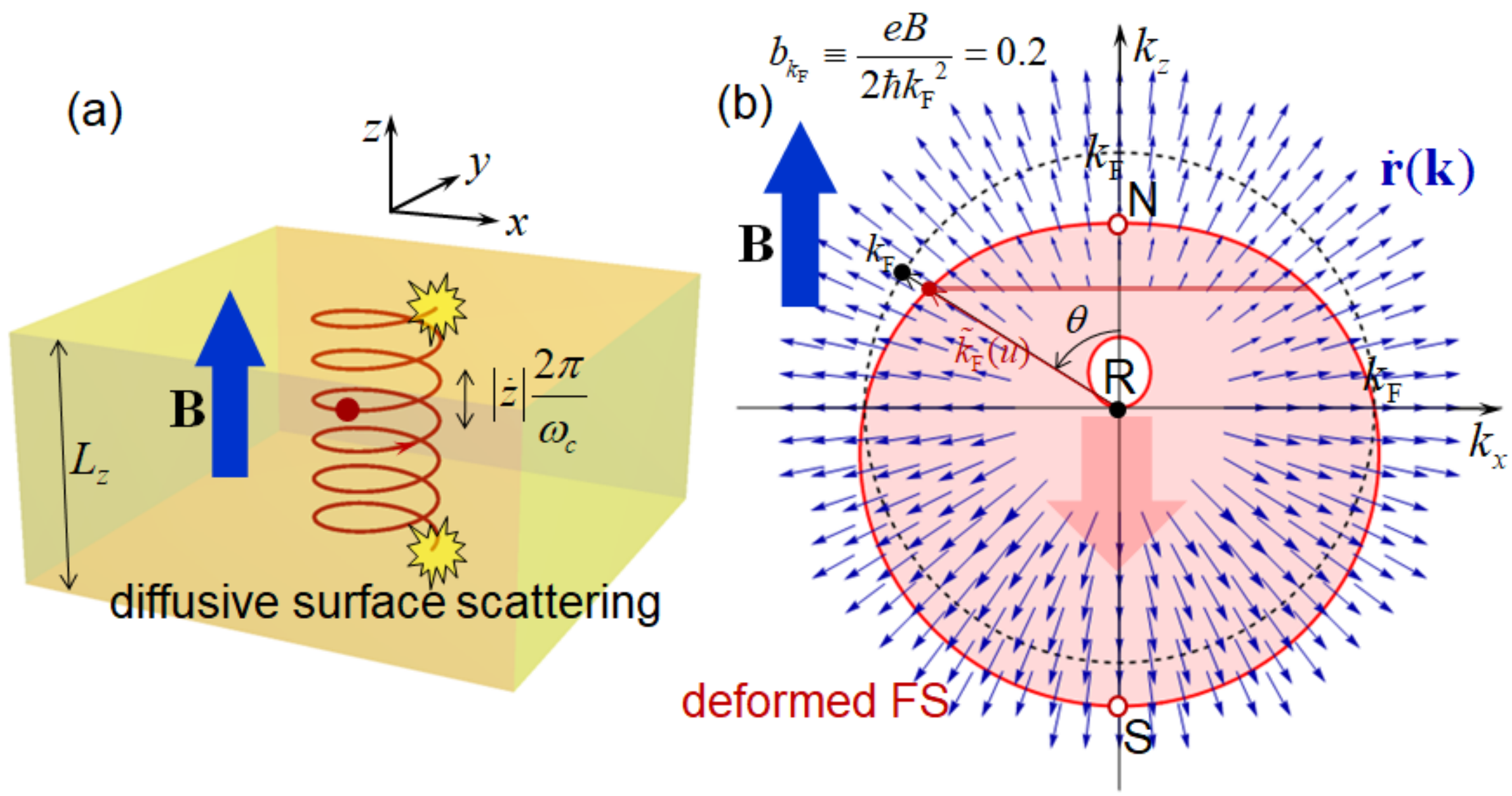

**FIG. 1.** (color online)

(a) Schematic illustration of the Sondheimer oscillation in a thin-film metal, arising from the commensurability between the film thickness and the pitch of the electron orbit. (b) Magnetic-field-induced deformation of the Fermi surface and the velocity field in the $k_x$-$k_z$ plane for the conduction band of a single Weyl cone with chirality $\chi$=+1. The two limiting points are denoted by N and S.

**Figure 2** (Osada)

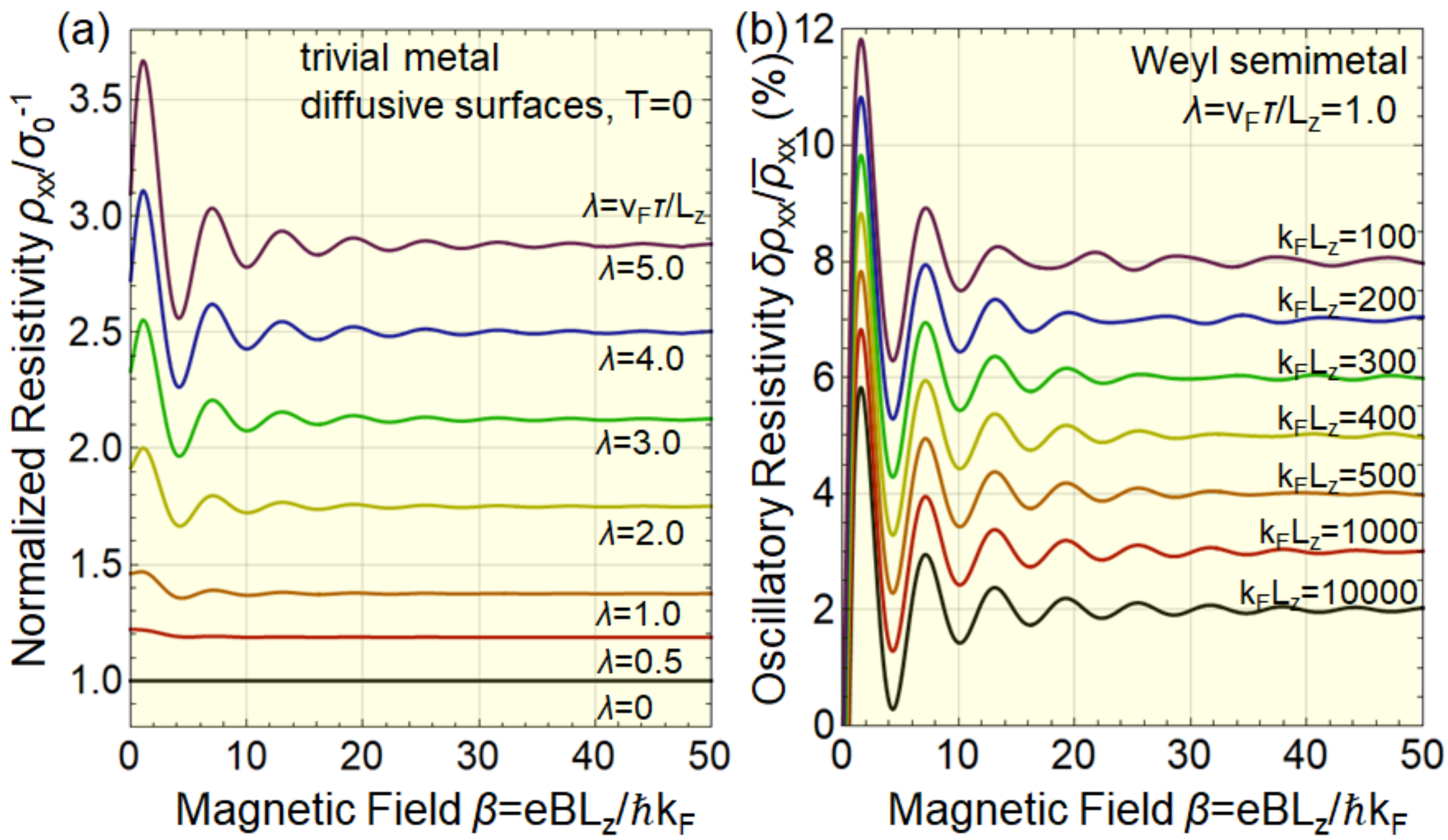

**FIG. 2.** (color online)

(a) Magnetic-field dependence of the in-plane resistivity $\rho_{xx}$ of a topologically trivial metal for several values of the size parameter $\lambda = v_F\tau / L_z$ . Magnetic-field-periodic Sondheimer oscillations are clearly seen. (b) Oscillatory component $\delta\rho_{xx}$ for a Weyl semimetal at fixed $\lambda = 1$ and several values of $k_F L_z$. The curves are vertically offset for clarity. The modulation becomes more pronounced as $k_F L_z$ decreases. The smallest-$k_F L_z$ curve is shown mainly to illustrate the trend.

**Figure 3** (Osada)

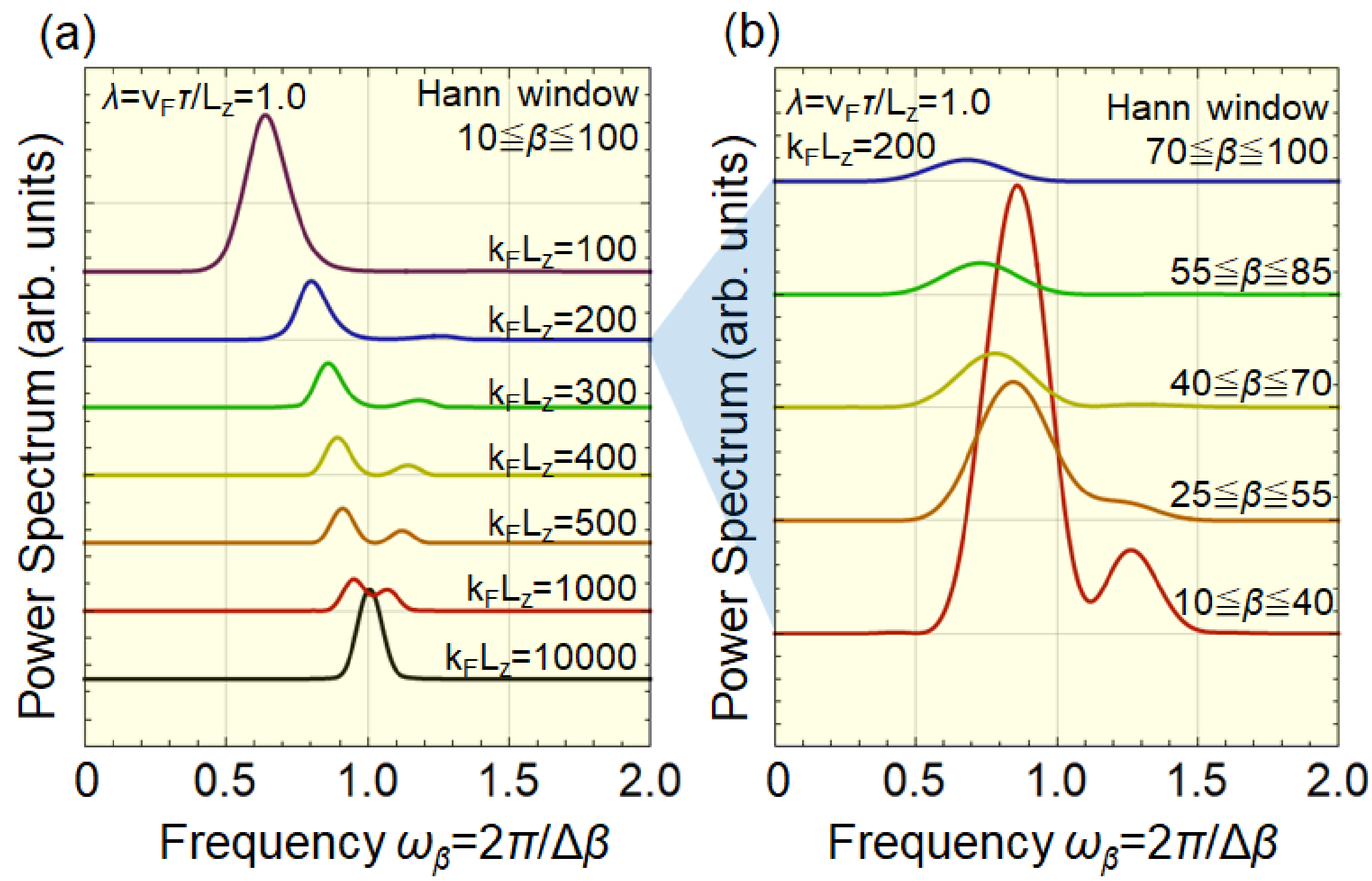

**FIG. 3.** (color online)

(a) Fourier power spectra of the oscillatory resistivity shown in Fig. 2(b), calculated over $10 \leqq \beta \leqq 100$ using a Hann window. The single peak for $k_F L_z$=10000 splits into two peaks with unequal intensities as $k_F L_z$ decreases. (b) Moving-window Fourier power spectra for $k_F L_z$ = 200. A Hann window is applied over $\beta_1 \leqq \beta \leqq \beta_2$. The separation between the two peaks increases as the window moves to higher $\beta$, demonstrating the chirp of the two SO branches.